\documentclass[aps,twocolumn,superscriptaddress,showpacs]{revtex4}
\usepackage{epsfig}
\usepackage{amsbsy}
\usepackage{amsmath}
\usepackage{amsfonts}

\begin{document}
\addtolength{\abovecaptionskip}{-2mm}
\addtolength{\belowcaptionskip}{-4mm}
\addtolength{\intextsep}{-6mm}
\addtolength{\abovedisplayskip}{-1mm}
\addtolength{\belowdisplayskip}{-1mm}
\title{\large Optimal States and Almost Optimal Adaptive Measurements for
Quantum Interferometry}
\author{D.\ W.\ Berry}
\affiliation{Department of Physics, The University of Queensland, St.\ Lucia
4072, Australia}
\author{H.\ M.\ Wiseman}
\affiliation{School of Science, Griffith University, Nathan, Brisbane,
Queensland 4111, Australia}
\date{\today}
\begin{abstract}
We derive the optimal $N$-photon two-mode input state for obtaining an estimate
$\phi$ of the phase difference between two arms of an interferometer.
For an optimal measurement [B.\ C.\ Sanders and G.\ J.\ Milburn, Phys.\ Rev.\ 
Lett.\ {\bf 75}, 2944 (1995)], it yields a variance $(\Delta \phi)^2
\simeq \pi^2/N^2$, compared to $O(N^{-1})$ or $O(N^{-1/2})$ for states
considered by previous authors.  Such a measurement cannot be
realized by counting photons in the interferometer outputs.  However,
we introduce an adaptive measurement scheme that can be thus realized,
and show that it yields a variance in $\phi$ very close to
that from an optimal measurement.
\end{abstract}
\maketitle

\newcommand{\nn}{\nonumber}
\newcommand{\nl}[1]{\nn \\ && {#1}\,}
\newcommand{\erf}[1]{Eq.\ (\ref{#1})}
\newcommand{\dg}{^\dagger}
\newcommand{\ip}[1]{\langle{#1}\rangle}
\newcommand{\bra}[1]{\langle{#1}|}
\newcommand{\ket}[1]{|{#1}\rangle}
\newcommand{\braket}[2]{\langle{#1}|{#2}\rangle}
\newcommand{\sq}[1]{\left[ {#1} \right]}
\newcommand{\cu}[1]{\left\{ {#1} \right\}}
\newcommand{\ro}[1]{\left( {#1} \right)}
\newcommand{\tr}[1]{{\rm Tr}\sq{ {#1} }}
\newcommand{\st}[1]{\left| {#1} \right|}

\begin{figure}[b]
\includegraphics[width=0.43\textwidth]{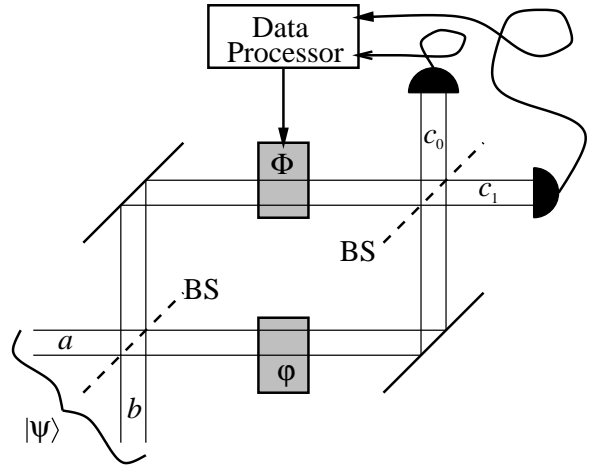}
\caption{The Mach-Zehnder interferometer, with the
addition of a controllable phase $\Phi$ in one arm. The unknown phase to be
estimated is $\varphi$. Both beam splitters (BS) are 50:50. }
\label{diag}
\end{figure}

Interferometry is the basis of many high-precision measurements. The ultimate
limit to the precision is due to quantum effects. This limit is most easily
explored for a Mach-Zehnder interferometer (see Fig.\ \ref{diag}, where $\Phi$
should be ignored for the moment). The outputs of this device can be measured to
yield an estimate $\phi$ of the phase difference $\varphi$ between the two arms
of the interferometer. It is well known that this can achieve the standard
quantum limit for phase sensitivity of $(\Delta \phi)^{2} \simeq N^{-1}$ when an
$N$-photon number state enters one input port. Several authors
\cite{Caves,Yurke,Holland,SandMil95,SandMil97} have proposed ways of reducing
the phase variance to the Heisenberg limit of ${\sim}N^{-2}$. Here $N$ is the
fixed total number of photons in the inputs \cite{fn1}.

Most of these proposals \cite{Caves,Yurke,Holland} are limited in that they
require that the phase difference $\varphi$ between the two arms be zero or
very small in order to obtain the $N^{-2}$ scaling. Sanders and Milburn
\cite{SandMil95,SandMil97} considered an ideal or canonical measurement, for
which the $N^{-2}$ scaling is independent of $\varphi$. Unfortunately they do
not explain how this measurement can be performed, and it can be shown
\cite{BerWis01} that it cannot be realized by counting photons in the outputs of
the interferometer. In this Letter we show that there is an experimentally
realizable measurement scheme using photodetectors and feedback which is almost
as good as the canonical measurement.

Before introducing our adaptive scheme, we find the optimal input states for the
canonical interferometric measurements. These will then be used as the input
states for our adaptive scheme, to demonstrate a $(\Delta \phi)^{2}$ scaling
almost as good as $N^{-2}$. The optimal input states are interesting in
themselves, in that they differ significantly from the input states considered
in Refs.\ \cite{Caves,Yurke,Holland,SandMil95,SandMil97}. In particular, our
rigorous analysis shows that those non-optimal states, in fact, exhibit a
{\em worse} scaling than the standard quantum limit of $N^{-1}$.

{\em The canonical measurement.}---Using the same notation as Sanders and
Milburn \cite{SandMil95}, we designate the two annihilation operators for the
two input modes as $\hat a$ and $\hat b$, and we use the Schwinger
representation
\begin{equation}
\hat J_x = (\hat a^\dagger \hat b + \hat a \hat b^\dagger )/2, ~~~
\hat J_y = (\hat a^\dagger \hat b - \hat a \hat b^\dagger )/2i,
\end{equation}
\begin{equation}
\hat J_z = (\hat a^\dagger \hat a - \hat b^\dagger \hat b )/2, ~~~
\hat J^2 = \hat J_x^2 + \hat J_y^2 + \hat J_z^2.
\end{equation}
We use the notation $\ket{j \mu}_z$ for the common eigenstate of $\hat J_z$
and $\hat J^2$ with eigenvalues $\mu$ and $j(j+1)$, respectively. This state
corresponds to Fock states with $j+\mu$ and $j-\mu$ photons entering ports
$a$ and $b$, respectively.

From Ref.\ \cite{SandMil97}, the optimal probability operator measure (POM) for
phase measurements is the canonical one,
\begin{equation}
\hat E(\phi) d\phi = \frac{2j+1}{2\pi} \ket{j\phi} \bra{j\phi} d\phi,
\end{equation}
where the phase states $\ket{j\phi}$ are defined in terms of the $\hat{J}_{y}$
eigenstates by
$\ket{j\phi}=(2j+1)^{-1/2} \sum_{\mu=-j}^j e^{i\mu \phi}\ket{j\mu}_y$.
In terms of the $\hat{J}_{y}$ eigenstates, the POM is
\begin{equation} \label{SMPOM}
\hat E(\phi) d\phi = \frac 1{2\pi} \sum_{\mu,\nu=-j}^j e^{i(\mu-\nu)\phi}
\ket{j\mu}_y \bra{j\nu} d\phi.
\end{equation}
The POM defines the probability distribution for $\phi$, the best estimate for
the interferometer phase $\varphi$, by
\begin{equation} \label{SMPM}
P(\phi)d\phi = \bra{\psi}\hat{E}(\phi)\ket{\psi}d\phi.
\end{equation}
Here $\ket{\psi}$ is the two-mode interferometer input state having $N=2j$
photons.

{\em The optimal input state.}---A short examination reveals that the canonical
POM (\ref{SMPOM}) has the same matrix elements as the POM for ideal measurements
on a single mode with an upper limit of $N$ on the photon number. The optimal
state in this case has been considered before \cite{SumPeg90,semiclass}. Here we
follow the procedure of Ref.\ \cite{semiclass}, which minimizes the Holevo phase
variance
\cite{Hol84}
\begin{equation} \label{Holevo}
(\Delta\phi)^{2} = V_{\phi} \equiv S_{\phi}^{-2}-1,
\end{equation}
where $S_{\phi} \in [0,1]$ is the {\em sharpness} of the phase distribution,
defined as
\begin{equation}
S_{\phi} \equiv \st{\ip{e^{i\phi}}} =  \int_{0}^{2\pi}
d\phi P(\phi) e^{i(\phi-\bar{\phi})},
\end{equation}
where the ``mean phase'' $\bar{\phi}$ is defined so that $S_{\phi}$ is
positive.
The Holevo variance is the natural quantifier for dispersion in a cyclic
variable. If the variance is small, then it is easy to show that
\begin{equation} \label{approxHol}
V_{\phi} \simeq \int_{0}^{2\pi}
4 \sin^{2}\ro{\frac{\phi-\bar{\phi}}{2}} P(\phi)d\phi ,
\end{equation}
from which the equivalence to the usual definition of variance for
well-localized distributions is readily apparent.

Using the Holevo variance enables a simple analytic solution.
The minimum variance is
\begin{equation}
 \tan^{2} \left( \frac \pi {N+2} \right)  = \frac{\pi^2}
{(N+2)^2} + O(N^{-4}),
\end{equation}
and the optimal state (chosen here to have a mean relative phase of zero) is
\begin{equation}
\ket{\psi_{\rm opt}}=\frac 1{\sqrt{j+1}} \sum_{\mu=-j}^{j} \sin \left[
\frac{(\mu+j+1)\pi} {2j+2}\right] \ket{j\mu}_{y}.
\end{equation}

To obtain the state in terms of the eigenstates of $\hat J_z$, we use
$_y \braket{j \mu}{j \nu}_z=e^{i(\pi /2)(\nu-\mu)}I_{\mu \nu}^j (\pi/2)$,
where \cite{SandMil95}
\begin{align}
I_{\mu \nu}^j (\pi /2) &= 2^{-\mu}\left[ \frac{(j-\mu)!}{(j-\nu)!}
\frac{(j+\mu)!}{(j+\nu)!} \right]^{1/2}
P_{j-\mu}^{(\mu-\nu,\mu+\nu)}(0),\nn \\
&\textrm{for } \mu-\nu>-1,\;\;  \mu+\nu>-1,
\end{align}
where $P_n^{(\alpha,\beta)}(x)$ are the Jacobi polynomials, and the other
matrix elements are obtained using the symmetry relations
$I_{\mu \nu}^j (\theta)=(-1)^{\mu-\nu}I_{\nu \mu}^j
(\theta)=I_{-\nu,-\mu}^j (\theta)$.

\begin{figure}[t]
\includegraphics[width=0.415\textwidth]{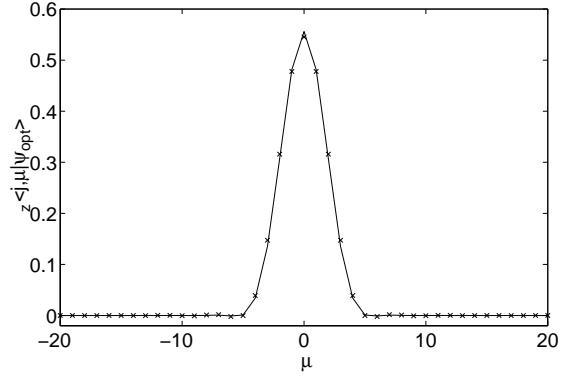}
\caption{The coefficients ${}_{z}\bra{j\mu}\psi_{\rm opt}\rangle$ for the state
optimized for minimum phase variance under ideal measurements. All coefficients
for a photon number of $2j=40$ are shown as the continuous line, and those near
$\mu=0$ for a photon number of $2j=1200$ as crosses.}
\label{input}
\end{figure}

An example of the optimal state for 40 photons is plotted in Fig.\ \ref{input}.
This state contains contributions from all the $\hat J_z$ eigenstates,
but the only significant contributions are from 9 or 10 states near
$\mu=0$. The distribution near the center is fairly independent of photon
number $N=2j$. To demonstrate this, the distribution near the center for
1200 photons is also shown in Fig.\ \ref{input}.

\begin{figure}[b]
\includegraphics[width=0.415\textwidth]{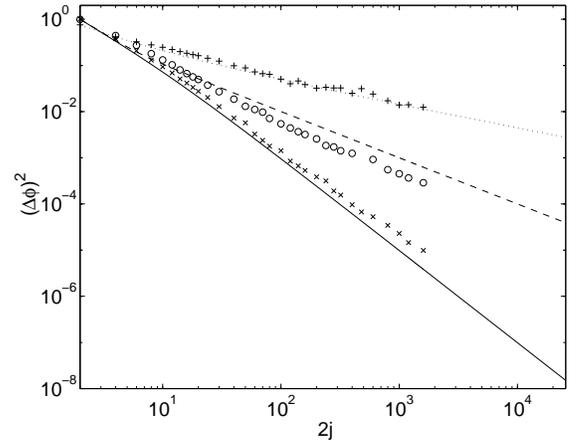}
\caption{Variances in the phase estimate versus input photon number $2j$. The
lines are exact results for canonical measurements on optimal states
$\ket{\psi_{\rm opt}}$ (continuous line), on states with all photons incident
on one input port $\ket{jj}_{z}$ (dashed line), and on states with equal photon
numbers incident on both input ports $\ket{j0}_{z}$ (dotted line). The crosses
are the numerical results for the adaptive phase measurement scheme on
$\ket{\psi_{\rm opt}}$, and the plusses are those on a $\ket{j0}_z$ input state.
The circles are numerical results for a nonadaptive phase measurement scheme on
$\ket{\psi_{\rm opt}}$. All variances for the $\ket{j0}_{z}$ state are for phase
modulo $\pi$.}
\label{three}
\end{figure}

In order to compare this state with $\ket{j0}_z$, where equal photon numbers
are fed into both input ports (as considered in
Refs.\ \cite{Holland,SandMil95,SandMil97}), the exact phase variance for this
case was calculated for a range of photon numbers up to 25$\,$600. Since the
phase must be measured modulo $\pi$ for this state, rather than using the Holevo
phase variance, we used the following measure for the dispersion:
\begin{equation} \label{modpi}
(\Delta\phi)^{2} = V_{2\phi}/4=(|\ip{e^{2i\phi}}|^{-2}-1)/4,
\end{equation}
where the expectation value is again determined using (\ref{SMPM}). The phase
variances for this state and the optimal state are shown in Fig.\ \ref{three}.
The exact Holevo phase variance of the state where all the photons are incident
on one port, $\ket{jj}_z$, is also shown for comparison.

As can be seen, the phase variance for $\ket{j0}_z$ scales down with photon
number much more slowly than the phase variance for optimal states \cite{fn2},
and even more slowly than the variance for $\ket{jj}_z$, which scales
as $N^{-1}$. In fact, this figure shows that the phase variance for
$\ket{j0}_z$ scales as $N^{-1/2}$, which agrees with what can be calculated
from the asymptotic formula for $P(\phi)$ given in Ref.\ \cite{SandMil97}. This
is a radically different result from the $N^{-2}$ scaling found in
Refs.\ \cite{Holland,SandMil95,SandMil97}. The state
$(\ket{j0}_{z}+\ket{j1}_{z})/\sqrt{2}$, considered in Ref.\ \cite{Yurke},
is even worse than the state $\ket{j0}_{z}$.

\begin{figure}[b]
\includegraphics[width=0.415\textwidth]{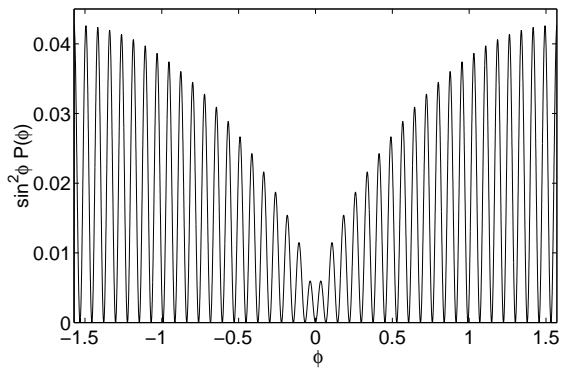}
\caption{The canonical phase probability distribution for $\ket{j0}_z$
multiplied by $\sin^{2}\phi$ for $2j=80$ photons.}
\label{dist}
\end{figure}

The reason for this discrepancy is that the results found in
Refs.\ \cite{Holland,SandMil95,SandMil97} are all based on the width of the
central peak in the distribution, but the main contribution to the variance is
from the {\it tails} of the distribution. This can be seen from the probability
distribution multiplied by $\sin^2\phi$, since Eqs.\ (\ref{approxHol}) and
(\ref{modpi}) imply that $(\Delta\phi)^{2} \simeq \int \sin^{2}(\phi-\varphi)
P(\phi) d\phi$. We plot this in Fig.\ \ref{dist} for $N=80$ and $\varphi=0$. In
practice this means that the error in the phase will be small most of the time,
but there will be a significant number of results with a large error.

{\em Adaptive measurements.}---Although the quantum interferometry problem is
now formally solved, this is of little practical use because (even if the
optimal input states could be produced) it is not known how to implement
the canonical measurement scheme. In particular, it is impossible to
implement it by counting photons in the two output ports of the interferometer
\cite{BerWis01}, as an experimenter would expect to do. Nevertheless, as we
will show, it is possible to closely approximate the canonical measurement
by counting output photons if one makes the measurement {\em adaptive}. The
situation is as in Fig.\ \ref{diag}. The unknown phase we wish to measure,
$\varphi$, is in one arm of the interferometer, and we introduce a known phase
shift, $\Phi$, into the other arm of the interferometer. After each
photodetection we adjust this introduced phase shift in order to minimize the
expected uncertainty of our best phase estimate after the next photodetection.

The annihilation operators $\hat c_{1}$ and $\hat c_{0}$ for the two
output modes shown in Fig.\ \ref{diag} are related to the inputs by
\begin{align} \label{anni}
\hat c_{u}(\varphi,\Phi) = \hat a \sin \frac{\varphi-\Phi + u\pi}{2} +
\hat b \cos \frac{\varphi-\Phi+u\pi}{2}. \nn \\
\end{align}
{\em Before} the $m$th photon has been detected, the phase $\Phi$ will
be fixed to the value $\Phi_{m}$ by an adaptive algorithm to be specified
later. Clearly the feedback loop which adjusts $\Phi_{m}$ must
act much faster than the average time between photon arrivals.
It is the ability to change $\Phi_{m}$ during the passage of a
single (two-mode) pulse that makes photon counting measurements
more general than a measurement of the  output $\hat{J}_{z}$ considered in
Refs.\ \cite{Yurke,Holland}.

An adjustable second phase $\Phi$ is of use even without feedback \cite{Hra96}.
By setting
\begin{equation} \label{nonadapt}
\Phi_m = \Phi_0 + \frac{m\pi}{N},
\end{equation}
where $\Phi_0$ is chosen randomly, an estimate $\phi$ of $\varphi$ can be made
with an accuracy independent of $\varphi$. However, as we will show, this phase
estimate has a variance scaling as $O(N^{-1})$ rather than $O(N^{-2})$.

Let us denote the result $u$ from the $m$th detection as $u_m$ (which is 0 or 1
according to whether the photon is detected in mode $c_{0}$ or $c_{1}$), and the
measurement record up to and including the $m$th detection as the binary string
$n_m\equiv u_m\cdots u_2u_1$. The input state after $m$ detections will be a
function of the measurement record and $\varphi$, and we denote it as
$\ket{\tilde\psi(n_m,\varphi)}$. It is determined by the initial condition
$\ket{\tilde\psi(n_0,\varphi)}=\ket{\psi}$ and the recurrence relation
\begin{equation} \label{rr}
\ket{\tilde\psi(u_{m}n_{m-1},\varphi)} = \hat
c_{u_{m}}(\varphi,\Phi_{m})
\frac{\ket{\tilde\psi(n_{m-1},\varphi)}}{\sqrt{N+1-m}}.
\end{equation}

These states are unnormalized, and their norm represents the probability for the
record $n_{m}$, given $\varphi$,
\begin{equation}
P(n_{m}|\varphi) = \braket{\tilde\psi(n_m,\varphi)}
{\tilde\psi(n_m,\varphi)}.
\end{equation}
Thus the probability of obtaining the result $u_m$ at the $m$th measurement,
given the previous results $n_{m-1}$, is
\begin{equation}
\label{resprob}
P(u_m|\varphi,n_{m-1})=\frac{\braket{\tilde\psi(u_m n_{m-1},\varphi)}
{\tilde\psi(u_m n_{m-1},\varphi)}}{\braket{\tilde\psi(n_{m-1},\varphi)}
{\tilde\psi(n_{m-1},\varphi)}}.
\end{equation}
Also, the posterior probability distribution for
$\varphi$ is
\begin{equation}
\label{probdist}
P(\varphi|n_m) = {\cal N}_{m}(n_{m})
\braket{\tilde\psi(n_m,\varphi)}{\tilde\psi(n_m,\varphi)},
\end{equation}
where ${\cal N}_m(n_{m})$ is a normalization factor. To obtain this we
have used Bayes' theorem assuming a flat prior distribution for
$\varphi$ (that is, an initially unknown phase). A Bayesian approach
to interferometry has been considered previously, and even realized
experimentally~\cite{Hra96}. However, this was done
only with nonadaptive measurements and with all particles
incident on one input port.

With this background, we can now specify the adaptive algorithm for
$\Phi_{m}$.
The sharpness of the distribution after the $m$th detection is given by
\begin{equation} \label{sharp}
S_{\varphi}(u_{m}n_{m-1})=\left| \int_{0}^{2\pi}
 P(\varphi|u_{m}n_{m-1}) e^{i\varphi}
d\varphi \right|.
\end{equation}
We choose the feedback phase before the $m$th detection, $\Phi_m$,
to maximize the sharpness. Since we do not know $u_m$ beforehand, we weight
the sharpnesses for the two alternative results by their probability of
occurring based on the previous measurement record. Therefore the expression
we maximize is
\begin{equation}
M(\Phi_{m}|n_{m}) = \sum_{u_{m}=0,1} P(u_{m}|n_{m-1})
S_{\varphi}(u_{m}n_{m-1}).
\end{equation}
Using Eqs.\ (\ref{resprob}), (\ref{probdist}), and (\ref{sharp}), and ignoring
the constant ${\cal N}_{m}(n_{m})$, the maximand can be rewritten as
\begin{equation}
\sum_{u_{m}=0,1}
\left|\int_0^{2\pi} \braket{\tilde\psi(u_{m}n_{m-1},\varphi)}
{\tilde\psi(u_{m}n_{m-1},\varphi)} e^{i\varphi} d \varphi  \right| .
\label{implic}
\end{equation}
The controlled phase $\Phi_{m}$ appears implicitly in (\ref{implic}) through
$\hat{c}_{u_{m}}(\varphi,\Phi_{m})$ in (\ref{anni}), which appears in the
recurrence relation (\ref{rr}). The maximizing solution $\Phi_{m}$ can be found
analytically, but we will not exhibit it here.

The final part of the adaptive scheme is choosing the phase
estimate $\phi$ of $\varphi$ from the complete data set $n_{N}$. To
maximize ${\rm Re}[\ip{ e^{i(\phi-\varphi)} }]$ (that is, to
minimize the deviation of $\phi$ from $\varphi$), we choose
$\phi$ to be the mean of the posterior distribution
$P(\varphi|n_{N})$, which from \erf{probdist} is
\begin{equation}
\phi = \arg\int_{0}^{2\pi} \braket{\tilde\psi(n_{N},\varphi)}
{\tilde\psi(n_{N},\varphi)} e^{i\varphi} d\varphi.
\end{equation}

We can determine the approximate phase variance $(\Delta \phi)^{2}$ under this
adaptive scheme using a stochastic method. The phase $\varphi$ is taken to be
zero, which leads to no loss of generality since the initial controlled phase
$\Phi_{1}$ is chosen randomly. The measurement results are determined randomly
with probabilities determined using $\varphi=0$, and the final estimate $\phi$
determined as above. From \erf{Holevo}, an ensemble
$\{ \phi_{\mu}\}_{\mu=1}^{M}$ of $M$ final estimates allows the phase variance
to be approximated by
$(\Delta\phi)^{2} \simeq -1+|{M}^{-1}\sum_{\mu=1}^{M}e^{i\phi_{\mu}}|^{-2}$.
It is also possible to determine the phase variance exactly by
systematically going through all the possible measurement records and
averaging over $\Phi_{1}$. This method is feasible only for photon
numbers up to 20 or 30, however.

The results of using this adaptive phase measurement scheme on
the optimal input states determined above are shown in Fig.\ \ref{three}.
The phase variance is very close to the phase variance for ideal
measurements, with scaling very close to $N^{-2}$.
The phase variances do differ relatively more from the ideal values for larger
photon numbers, however, indicating a scaling slightly worse than $N^{-2}$. It
is possible that the actual scaling is $\log N/N^2$, as is the case for
optimal single-mode adaptive phase measurements based on
dyne detection \cite{fullquan,unpub}. For comparison,
we also show the variance from the nonadaptive phase measurement defined by
\erf{nonadapt}. As is apparent, this has a variance scaling as $N^{-1}$.
Finally, the results for $\ket{j0}_z$ input states with this adaptive phase
measurement scheme (modified to estimate $\varphi$ modulo $\pi$, as we must for
these states) are also shown. The variances are again close to those for ideal
measurements, scaling as $N^{-1/2}$.

To conclude, we have shown that although the $\ket{j0}_z$
input states considered by previous authors do not give
a phase variance that scales as $1/N^2$ under ideal measurements, it is
straightforward to derive optimal input states that do give this
scaling. These states require significant
contributions from only 9 or 10 $\hat J_z$ eigenstates (photon
eigenstates for the two input modes). We have also demonstrated a practical
measurement scheme (i.e., one based on photodetection of the output modes) to
approximate ideal measurements. This scheme, which uses feedback,
allows phase measurements with a variance scaling close to $1/N^2$
with optimized input states.


\begin{thebibliography}{99}
\bibitem{fn1}
Except for Ref.\ \cite{Caves}, the above proposals have involved input states
where the total number of photons  is fixed. For simplicity, we follow this
convention.
\bibitem{BerWis01}
We have found that for $N=3$ and $N=4$ our feedback scheme as applied to
optimum input states is globally optimum, and also that the phase variance is
above the canonical phase variance.  This demonstrates that canonical
measurements cannot be realized by counting photons in the interferometer
outputs, even if feedback is allowed.
\bibitem{fn2} The phase variance
starts out less than that for the optimum state, but this is merely due to
using a slightly different measure of the variance.
\end{thebibliography}
\end{document}